\documentclass{iopart}

\usepackage{graphicx}
\usepackage{iopams}

\newcommand{\bra}[1]{\langle #1|}
\newcommand{\ket}[1]{|#1\rangle}

\begin{document}

\title
{Transport through two-level quantum dots weakly coupled to ferromagnetic leads}

\author{I. Weymann$^1$ and J. Barna\'s$^{1,2}$}
\address{$^1$ Department of Physics, Adam
Mickiewicz University, 61-614 Pozna\'n, Poland}
\address{$^2$ Institute of Molecular Physics, Polish
Academy of Sciences, 60-179 Pozna\'n, Poland}
\ead{weymann@amu.edu.pl}

\date{\today}

\begin{abstract}
Spin-dependent transport through a two-level quantum dot in the
sequential tunneling regime is analyzed theoretically by means of
a real-time diagrammatic technique. It is shown that the current,
tunnel magnetoresistance, and shot noise (Fano factor) strongly
depend on the transport regime, providing a detailed information
on the electronic structure of quantum dots and their coupling to
external leads. When the dot is asymmetrically coupled to the
leads, a negative differential conductance may occur in certain
bias regions, which is associated with a super-Poissonian shot
noise. In the case of a quantum dot coupled to one half-metallic
and one nonmagnetic lead, one finds characteristic Pauli spin
blockade effects. Transport may be also suppressed when the dot
levels are coupled to the leads with different coupling strengths.
The influence of an external magnetic field on transport
properties is also discussed.
\end{abstract}


\section{Introduction}

Quantum dots have already paved their way to become underlying
devices of magnetoelectronics and spintronics
\cite{wolf01,loss02,maekawa02,zutic04} -- not only because of
beautiful physics emerging in those systems, but, more
importantly, due to possible future applications and due to the
possibility of manipulation of a single spin
\cite{kouwenhoven97,kouwenhoven98,ralph02,hanson03}. Transport
characteristics of quantum dots coupled to nonmagnetic leads have
already been extensively studied both theoretically and
experimentally. In nonmagnetic single-electron devices it is
mainly the electron charge which determines the system transport
properties \cite{nato92,sohn97,likharev99,heinzel03}. If, however,
a quantum dot is coupled to ferromagnetic leads, transport
characteristics strongly depend on the spin degree of freedom,
leading for example to the suppression of current when the
alignment of magnetic moments of the leads switches from parallel
to antiparallel \cite{julliere75,barnas98,rudzinski01}. The
difference between the currents flowing through the system in
these two magnetic configurations defines the so-called tunnel
magnetoresistance (TMR).

Theoretical considerations of electronic transport through a
quantum dot weakly coupled to ferromagnetic leads were restricted
mainly to single-level quantum dots. Transport properties of such
systems were analyzed in the sequential tunneling regime
\cite{rudzinski01,bulka00,koenigPRL03}, as well as in the
cotunneling regime
\cite{weymannPRB05,weymannPRBBR05,weymannEPJ05}. In real systems,
however, usually more than one energy level participate in
transport, leading to more complex and interesting transport
characteristics \cite{thielmann05,belzig04,elste06}. Several
experimental realizations of quantum dots attached to
ferromagnetic contacts have already been reported, which include
self-assembled dots in ferromagnetic semiconductors \cite{chye02},
ultrasmall metallic grains \cite{ralph02,heersche06}, granular
structures \cite{zhang05}, carbon nanotubes
\cite{tsukagoshi99,zhao02,jensen05,sahoo05}, single molecules
\cite{pasupathy04}, or magnetic tunnel junctions \cite{fertAPL06}.

In this paper we consider transport through two-level quantum dots
coupled to ferromagnetic leads, and restrict our considerations to
the sequential tunneling regime. The dot is described by the
Anderson-like impurity Hamiltonian. Thus, the analysis may also
hold for some magnetic impurities and molecules. With the aid of
the real-time diagrammatic technique, we calculate the current
$I$, differential conductance $G$, and Fano factor $F$ in the
parallel and antiparallel magnetic configurations, as well as the
corresponding TMR. The Fano factor, $F=S/S_p$, describes the
deviation of the zero-frequency shot noise $S$ from the Poissonian
shot noise $S_p=2e|I|$ (corresponding to uncorrelated electronic
transport).

In the following we analyze transport properties in the two
situations: (i) when the dot is symmetrically coupled to the
leads, and (ii) when the dot is coupled asymmetrically to the
leads. In both cases transport characteristics are shown to be
strongly dependent on the transport regime. As a result, variation
of TMR with the transport and gate voltages is significantly
different from that for single-level quantum dots. Furthermore,
the Fano factor is found to exhibit a nontrivial dependence on the
magnetic configuration of the system and spin polarization of the
leads. For symmetric coupling to ferromagnetic leads, the Fano
factor becomes divergent in the parallel configuration when the
leads' polarization tends to unity, while in the antiparallel
configuration shot noise is Poissonian. On the other hand, in the
case of asymmetric coupling of the dot to external leads, we find
transport regions where current is (partly) suppressed. These
transport regions are accompanied by NDC and a super-Poissonian
shot noise. If the dot is coupled to one half-metallic and one
nonmagnetic lead, the blockade regions are associated with
particular occupation of a dot spin state and can be referred to
as the Pauli spin blockade regions. In addition, we also discuss
the effect of an external magnetic field on the transport
characteristics. The TMR displays then a rather complex behavior
with the bias voltage.

We note that such effects as spin blockade, NDC, sub- and
super-Poissonian shot noise, and tunnel magnetoresistance occur
also in single-level quantum dots \cite{rudzinski01,bulka00}.
Although the physical mechanisms responsible for some of the above
effects in two-level quantum dots are similar to those in
single-level dots, the resulting transport characteristics in
two-level dots are much more complex. The analysis of these
effects in the case of two-level quantum dots coupled to
ferromagnetic leads is the main objective of the present work.

The paper is organized as follows. In section 2 we describe the
model. Method used to calculate transport characteristics is
presented briefly in section 3. Results of numerical calculations
are shown and discussed in section 4, where we first consider the
case of a quantum dot coupled symmetrically to the leads, and then
analyze transport through quantum dots with unequal couplings.
Section 4 also includes some approximate analytical expressions
obtained for the most characteristic transport regions. The
corresponding formulas may be useful in discussing future
experiments. Finally, summary and conclusions are given in section
5.

\section{Model}

\begin{figure}[b]
\centering
  \includegraphics[width=0.55\columnwidth]{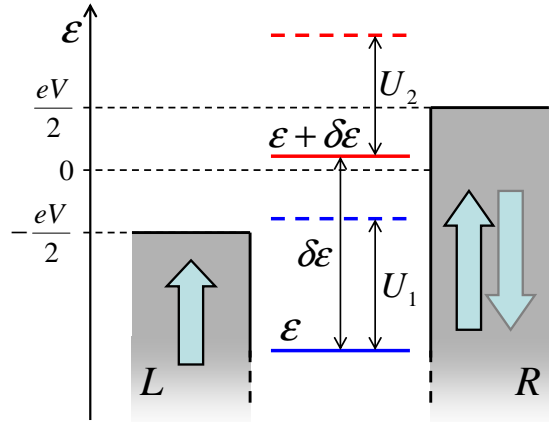}
  \caption{\label{Fig:1} (color online) Energy diagram of a two-level
  quantum dot coupled to ferromagnetic leads.
  For clarity reasons the energy diagram is shown here for $\Delta =U^\prime
  =0$. However, in the paper we consider the case when the Coulomb interaction between
  electrons occupying different orbitals is described by a nonzero value of
  the parameter $U^\prime$.
  The leads' magnetizations can form either
  parallel or antiparallel configurations.
  The arrows indicate the net spin of the leads.}
\end{figure}

The system considered in this paper is shown schematically in
Fig.~\ref{Fig:1} and consists of a quantum dot with two orbital
levels coupled to ferromagnetic leads. The net spin moments of the
leads are assumed to be collinear, i.e., they can form either
parallel or antiparallel magnetic configuration, as indicated in
Fig.~\ref{Fig:1}.
The Hamiltonian $\hat{H}$ of the system includes four terms,
$\hat{H}=\hat{H}_{\rm L} + \hat{H}_{\rm R} + \hat{H}_{\rm D} +
\hat{H}_{\rm T}$. The first two terms describe noninteracting
itinerant electrons in the leads, $\hat{H}_r=\sum_{{\mathbf
k}\sigma} \varepsilon_{r{\mathbf k}\sigma} c^{\dagger}_{r{\mathbf
k}\sigma} c_{r{\mathbf k}\sigma}$ for the left ($r={\rm L}$) and
right ($r={\rm R}$) leads, with $\varepsilon_{r{\mathbf k}\sigma}$
being the energy of an electron with the wave vector ${\mathbf k}$
and spin $\sigma$ in the lead $r$, and $c^{\dagger}_{r{\mathbf
k}\sigma}$ ($c_{r{\mathbf k}\sigma}$) denoting the respective
creation (annihilation) operator. The quantum dot is described by
the following Anderson-like Hamiltonian:
\begin{eqnarray}\label{Eq:Hamiltonian}\fl
  \hat{H}_{\rm D} =\sum_{j\sigma} \varepsilon_{j} n_{j\sigma}
  + \sum_j U_j n_{j\uparrow} n_{j\downarrow}
  + U^\prime \sum_{\sigma\sigma^\prime}
  n_{1\sigma}n_{2\sigma^\prime}
  -\frac{\Delta}{2} \sum_j
  \left(n_{j\uparrow}-n_{j\downarrow}\right) \,,
\end{eqnarray}
where $d^{\dagger}_{j\sigma}$ ($d_{j\sigma}$) is the creation
(annihilation) operator of an electron with spin $\sigma$ on the
$j$th level ($j=1,2$), $\varepsilon_{j}$ is the corresponding
single-particle energy, and $n_{j\sigma}$ is the particle number
operator, $n_{j\sigma} = d^{\dagger}_{j\sigma}d_{j\sigma}$. The
on-level Coulomb repulsion between two electrons of opposite spins
is described by $U_j$, whereas the inter-level repulsion energy is
denoted by $U^\prime$. The forth term in
Eq.~(\ref{Eq:Hamiltonian}) describes the Zeeman energy, with
$\Delta=g\mu_B B$ being the Zeeman splitting of the energy levels
($B$ is an external magnetic field along the magnetic moment of
the left electrode). To describe the energy structure of the dot
we introduce the level spacing $\delta\varepsilon= \varepsilon_2
-\varepsilon_1$ and define $\varepsilon_1\equiv \varepsilon$.
Since the different orbital levels can couple differently to an
applied magnetic field, $\delta\varepsilon$ can be controlled by
changing the magnetic field \cite{sasaki00}, while $\varepsilon$
can be tuned by a gate voltage.

Tunneling processes between the dot and electrodes are described
by the Hamiltonian,
\begin{equation}
  \hat{H}_{\rm T}=\sum_{r=\rm
  L,R}\sum_{{\mathbf k}j\sigma}\left(t_{rj}c^{\dagger}_{r {\mathbf k}\sigma}
  d_{j\sigma}+ t_{rj}^\star d^\dagger_{j\sigma} c_{r {\mathbf k}\sigma}
  \right) \,,
\end{equation}
where $t_{rj}$ denotes the tunnel matrix elements between the lead
$r$ and the $j$th dot level. Coupling of the $j$th level to
external leads can be described by $\Gamma_{rj}^{\sigma}= 2\pi
|t_{rj}|^2 \rho_r^\sigma$, with $\rho_r^\sigma$ being the
spin-dependent density of states in the lead $r$. By introducing
the spin polarization of the lead $r$, $p_{r}=(\rho_{r}^{+}-
\rho_{r}^{-})/ (\rho_{r}^{+}+ \rho_{r}^{-})$, the coupling
parameters $\Gamma_{rj}^{\sigma}$ can be expressed as
$\Gamma_{rj}^{+(-)}=\Gamma_{rj}(1\pm p_{r})$, with $\Gamma_{rj}=
(\Gamma_{rj}^{+} +\Gamma_{rj}^{-})/2$. Here, $\Gamma_{rj}^{+}$ and
$\Gamma_{rj}^{-}$ describe the coupling of the $j$th level to the
spin-majority and spin-minority electron bands, respectively. In
general, each dot level may be coupled to the leads with a
different strength. Moreover, the coupling strengths may be energy
dependent. In this work, however, they are assumed to be constant
within the electron bands. For example, in the case of experiments
performed in the weak coupling regime by Kogan {\it et al}
\cite{kogan04}, the coupling strength was found to be of the order
of tens of $\mu$eV.

\section{Method}

We analyze the spin-polarized sequential transport through a
two-level quantum dot. The first-order tunneling gives the
dominant contribution to charge current for voltages above a
certain threshold voltage and is exponentially suppressed in the
Coulomb blockade regime. The effects due to higher-order
tunneling, e.g., cotunneling \cite{weymannPRB05,cotunneling} are
not included. Taking into account the two orbital levels of the
dot results in sixteen different dot states $\ket{\chi}$
($\ket{\chi}=\ket{\chi_1}\ket{\chi_2}$, with $\ket{\chi_j}$
corresponding to the $j$-th level), as listed in
Table~\ref{tab:states}. The system is symmetrically biased and we
assume equal capacitive couplings to the left and right lead, so
the dependence of the dot energy levels on the bias voltage may be
neglected. The energies of the corresponding dot eigenstates are
listed in Table~\ref{tab:states}.

\begin{table}[t]
\caption{\label{tab:states}The dot eigenstates $\ket{\chi}$ and
their energies $\varepsilon_\chi$.} \centering
\begin{tabular}{lcc}
$\chi$ & Eigenstates & Energies\\
\hline
1 & $\ket{0}\ket{0}$ & 0\\
2 & $\ket{\uparrow}\ket{0}$ & $\varepsilon-\Delta/2$\\
3 & $\ket{\downarrow}\ket{0}$ & $\varepsilon+\Delta/2$\\
4 & $\ket{0}\ket{\uparrow}$ & $\varepsilon+\delta\varepsilon-\Delta/2$\\
5 & $\ket{0}\ket{\downarrow}$ & $\varepsilon+\delta\varepsilon+\Delta/2$\\
6 & $\ket{\uparrow}\ket{\downarrow}$ & $2\varepsilon+\delta\varepsilon+U^\prime$\\
7 & $\ket{\downarrow}\ket{\uparrow}$ & $2\varepsilon+\delta\varepsilon+U^\prime$\\
8 & $\ket{\uparrow}\ket{\uparrow}$ & $2\varepsilon+\delta\varepsilon+U^\prime-\Delta$\\
9 & $\ket{\downarrow}\ket{\downarrow}$ & $2\varepsilon+\delta\varepsilon+U^\prime+\Delta$\\
10 & $\ket{\uparrow\downarrow}\ket{0}$ & $2\varepsilon+U_1$\\
11 & $\ket{0}\ket{\uparrow\downarrow}$ & $2\varepsilon+2\delta\varepsilon+U_2$\\
12 & $\ket{\uparrow}\ket{\uparrow\downarrow}$ & $3\varepsilon+2\delta\varepsilon+U_2+2U^\prime-\Delta/2$\\
13 & $\ket{\downarrow}\ket{\uparrow\downarrow}$ & $3\varepsilon+2\delta\varepsilon+U_2+2U^\prime+\Delta/2$\\
14 & $\ket{\uparrow\downarrow}\ket{\uparrow}$ & $3\varepsilon+\delta\varepsilon+U_1+2U^\prime-\Delta/2$\\
15 & $\ket{\uparrow\downarrow}\ket{\downarrow}$ & $3\varepsilon+\delta\varepsilon+U_1+2U^\prime+\Delta/2$\\
16 & $\ket{\uparrow\downarrow}\ket{\uparrow\downarrow}$ &
$4\varepsilon+2\delta\varepsilon+U_1+U_2+4U^\prime$
\end{tabular}
\end{table}

Transport is calculated with the aid of the real-time diagrammatic
technique \cite{diagrams,thielmann,weymannPRB05,thielmann05}. The
technique consists in a systematic perturbation expansion of the
dot density matrix and the current operator in the dot-lead
couplings $\Gamma_{rj}$. In order to calculate the stationary
occupation probabilities, current, and shot noise in the lowest
order in the dot-lead coupling, we follow the procedure developed
in Refs. \cite{thielmann,thielmann05}, and introduce the
respective self-energy matrices: $\mathbf{W}$, $\mathbf{W^I}$,
$\mathbf{W^{II}}$. The matrix $\mathbf{W}$ contains self-energies
with one arbitrary row $\chi_0$ replaced by
$(\Gamma,\dots,\Gamma)$, which is due to the normalization of the
probabilities, $\sum_{\chi}P_{\chi}=1$. The elements $W_{\chi
\chi^\prime}$ of the matrix $\mathbf{W}$ describe the first-order
tunneling transitions between the many-body dot states
$\ket{\chi}$ and $\ket{\chi^\prime}$. They are given by
\cite{thielmann05}, $W_{\chi \chi^\prime}=W_{\chi \chi^\prime}
^{\rm L} + W_{\chi \chi^\prime}^{\rm R}$, where $W_{\chi
\chi^\prime}^r = 2\pi \sum_\sigma \rho_r^\sigma \left[
f_r(\varepsilon_\chi - \varepsilon_{\chi^\prime}) \left|\sum_j
t_{rj}^\star \bra{\chi}d_{j\sigma}^\dagger
\ket{\chi^\prime}\right|^2 + [1-f_r(\varepsilon_{\chi^\prime} -
\varepsilon_{\chi})] \left|\sum_j t_{rj} \bra{\chi}d_{j\sigma}
\ket{\chi^\prime}\right|^2 \right]$ for $\chi\neq\chi^\prime$,
while $W_{\chi \chi}^r = - \sum_{\chi^\prime\neq\chi}
W_{\chi^\prime\chi}^r$, with $f_r(\varepsilon) =
1/[e^{(\varepsilon-\mu_r)/k_{\rm B}T}+1]$ and $\mu_r$ being the
electrochemical potential of lead $r$. The second matrix,
$\mathbf{W^I}$, denotes the full self-energy matrix with one {\it
external} vertex, resulting from the expansion of the tunneling
Hamiltonian, replaced by the current operator. Finally, the third
matrix, $\mathbf{W^{II}}$, consists of self-energies with two {\it
external} vertices replaced by the current operator. The current
operator $\hat{I}$ is defined as $\hat{I}=(\hat{I}_{\rm R} -
\hat{I}_{\rm L})/2$, with $\hat{I}_r=-i(e/\hbar) \sum_{{\mathbf
k}\sigma} \sum_{j}\left(t_{rj}c^{\dagger}_{r {\mathbf k}\sigma}
d_{j\sigma}- t_{rj}^\star d^\dagger_{j\sigma} c_{r {\mathbf
k}\sigma}\right)$ being the current flowing from the lead $r$ to
the dot. The elements of the matrices $\mathbf{W^{I}}$ and
$\mathbf{W^{II}}$ can be expressed in terms of $W_{\chi
\chi^\prime}$ as \cite{thielmann05}, $W_{\chi \chi^\prime}^{\rm I}
= \left[ \Theta(N_{\chi^\prime}-N_{\chi})
-\Theta(N_{\chi}-N_{\chi^\prime}) \right] \left( W_{\chi
\chi^\prime}^{\rm R} - W_{\chi \chi^\prime}^{\rm L} \right)$ and
$W_{\chi \chi^\prime}^{\rm II} = (1-2 \delta_{\chi\chi^\prime})
W_{\chi \chi^\prime}/4$, respectively, where $N_\chi =
\sum_{j\sigma} n_{j\sigma}$ and $\Theta(x)$ is the step function.

In this paper we calculate transport in the first-order with
respect to tunneling processes. In the collinear configurations
discussed here, one might also apply a simpler method based on the
master equations. However, we apply a more sophisticated
technique, which enables easy extensions to noncollinear magnetic
configurations and facilitates including the influence of second-
and higher-order tunneling processes \cite{thielmann06}.
Furthermore, the real-time diagrammatic technique employed here
allows one to calculate the current and shot noise in a fully
systematic way order by order in the dot-lead coupling strength.

Having calculated the respective matrices, the stationary
probabilities can be determined from the following equation,
\begin{equation}\label{Eq:master}
  \left(\mathbf{W}\mathbf{P}\right)_{\chi}=
  \Gamma\delta_{\chi\chi_0}\,,
\end{equation}
where $\mathbf{P}$ is the vector containing the occupation
probabilities. In turn, the sequential current flowing through the
system can be calculated from
\begin{equation}\label{Eq:current}
  I=\frac{e}{2\hbar}{\rm Tr}\{\mathbf{W^I}\mathbf{P}\} \,,
\end{equation}
with ${\rm Tr}\{\mathbf{A}\}$ denoting the trace of the matrix
$\mathbf{A}$. On the other hand, the zero-frequency shot noise,
$S=2\int_{-\infty}^0 dt\left(\langle
\hat{I}(t)\hat{I}(0)+\hat{I}(0)\hat{I}(t)\rangle-2 \langle
\hat{I}\rangle^2 \right)$, is given by
\begin{equation}\label{Eq:noise}
  S=\frac{e^2}{\hbar}{\rm Tr}\{
  \mathbf{W^{II}}\mathbf{P}
  +\mathbf{W^{I}}\mathbf{\tilde{P}}\mathbf{W^{I}}\mathbf{P}
  \} \,,
\end{equation}
where $\mathbf{\tilde{P}}$ is determined from the equation
$\mathbf{W} \mathbf{\tilde{P}}=\mathbf{Q}$, with
$Q_{\chi^\prime\chi} =\left(P_{\chi^\prime}- \delta_{\chi^\prime
\chi}\right) \left(1-\delta_{\chi^\prime \chi_0}\right)$.

Knowing the current $I$ and the zero-frequency shot noise $S$, one
can determine the Fano factor $F$, $F=S/(2e|I|)$. The Fano factor
describes the deviation of $S$ from the Poissonian shot noise
given by $S_p=2e|I|$.

\section{Results and discussion}

In this section we present numerical and analytical results on the
current $I$, TMR, and Fano factor $F$ for the two-level quantum
dot weakly coupled to the leads. The TMR effect is
phenomenologically defined as \cite{julliere75,weymannPRB05}
\begin{equation}
   {\rm TMR} = \frac{I_{\rm P} - I_{\rm AP}}{I_{\rm AP}} \,,
\end{equation}
where $I_{\rm P}$ ($I_{\rm AP}$) is the current flowing through
the system when magnetic moments of the leads are aligned in
parallel (antiparallel).

In our considerations we will distinguish between quantum dots
coupled symmetrically and asymmetrically to the leads. The
asymmetry may be introduced in various ways -- for example by
attaching the quantum dot to leads with different spin
polarizations (e.g., to one half-metallic and one nonmagnetic
lead). On the other hand, if the leads have the same spin
polarizations, their coupling to the dot levels may be different.
In the following we first consider the case of quantum dots
symmetrically coupled to the leads and then we shall proceed to
analyze transport through quantum dots asymmetrically coupled to
the leads. In order to simplify the following discussion of
numerical and analytical results, we set $U_1=U_2=U^\prime \equiv
U$.

\subsection{Quantum dots symmetrically coupled to ferromagnetic leads}

\begin{figure}[b]
\centering
  \includegraphics[width=0.9\columnwidth]{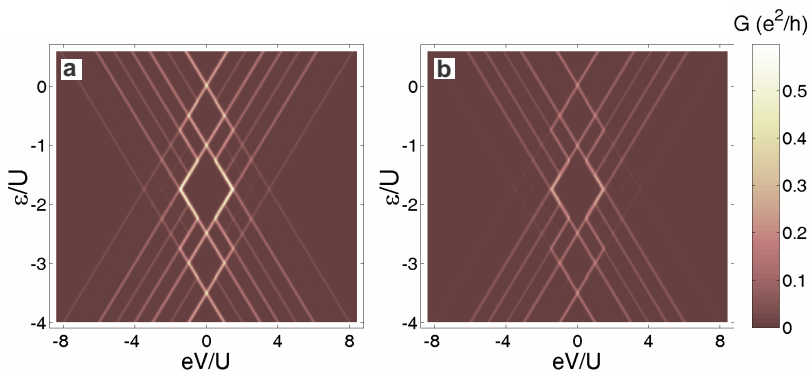}
  \caption{\label{Fig:2} (color online)
  The differential conductance $G=dI/dV$ as a function
  of the bias voltage and level position in
  the parallel (a) and antiparallel (b) magnetic configurations
  for the parameters:
  $k_{\rm B} T=\Gamma$, $\delta\varepsilon=25\Gamma$,
  $U=50\Gamma$, $\Delta=0$, $p_{\rm L} = p_{\rm R}\equiv p=0.7$, and
  $\Gamma_{rj}\equiv\Gamma/2$ ($r={\rm L,R}$, $j=1,2$).}
\end{figure}

In Fig.~\ref{Fig:2} we show the differential conductance as a
function of the bias voltage $V$ and level position $\varepsilon$.
Experimentally, the position of the dot level can be tuned by a
gate voltage, therefore Fig.~\ref{Fig:2} effectively shows the
bias and gate voltage dependence of the differential conductance.
The upper part [Fig.~\ref{Fig:2}(a)] corresponds to the parallel
magnetic configuration, whereas the lower part
[Fig.~\ref{Fig:2}(b)] shows the conductance in the antiparallel
configuration of the system. Both parts are plotted in the same
scale, which facilitates the comparison. First of all, one can
note that the conductance in antiparallel configuration is smaller
than that in the parallel one. This is due to the spin asymmetry
in tunneling processes, which leads to a partial suppression of
the conductance when the leads' magnetizations are antiparallel.
The majority (minority) electrons of the source electrode tunnel
then to the minority (majority) electron band of the drain
electrode, leading effectively to a conductance smaller than that
in the parallel configuration. If the spin asymmetry in the
density of states increases (the spin polarization of the leads
increases), the difference in conductance between these two
magnetic configurations becomes larger, which leads to an increase
in TMR.

Another feature visible in Fig.~\ref{Fig:2} is the diamond-like
structure of the differential conductance, which is associated
with discrete charging of the quantum dot. The diamonds around
$V=0$ correspond to the Coulomb blockade regions. When lowering
position of the dot levels, the charge of the dot changes
successively. More precisely, the dot is empty for
$\varepsilon\gtrsim 0$, occupied by one electron for
$0\gtrsim\varepsilon\gtrsim -U$, doubly occupied for
$-U\gtrsim\varepsilon\gtrsim -(2U+\delta\varepsilon)$, occupied by
three electrons for $-(2U+\delta\varepsilon)
\gtrsim\varepsilon\gtrsim -(3U+\delta\varepsilon)$, and the two
orbital levels of the dot are fully occupied for
$-(3U+\delta\varepsilon) \gtrsim\varepsilon$. In all these
transport regions the dot is in a well-defined charge state, and
the sequential tunneling is exponentially suppressed
\cite{kouwenhoven97,nato92}. If the bias voltage is increased
above a certain threshold voltage, the current starts to flow due
to first-order tunneling processes. When the charging energy of
the dot is much larger than the thermal energy, one observes then
a well-resolved step in the current as a function of the bias
voltage. In the density plots shown in Fig.~\ref{Fig:2}, this can
be seen in the form of lines that clearly separate the Coulomb
blockade regions from transport regions associated with
consecutive charge states taking part in transport. When the bias
voltage increases further, additional steps (and consequently
lines in Fig.~\ref{Fig:2}) arise at voltages where new states
becomes active in transport.

\begin{figure}[t]
\centering
  \includegraphics[width=0.45\columnwidth]{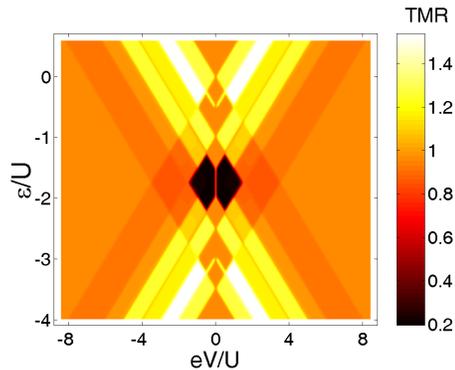}
  \caption{\label{Fig:3} (color online)
  The tunnel magnetoresistance as a function
  of the bias voltage and level position for
  the same  parameters as in Fig.~\ref{Fig:2}.}
\end{figure}

Figure~\ref{Fig:3} presents the corresponding TMR as a function of
the bias and gate voltages. By comparison with Fig.~\ref{Fig:2},
one can easily identify the regions of Coulomb blockade. First of
all, it is worth noting that, depending on the transport region,
TMR takes several well-defined values ranging roughly from $0.2$
to $1.6$, see Fig.~\ref{Fig:3}. Such behavior of TMR is
significantly different from that for a single-level quantum dot,
where  TMR in the sequential tunneling regime acquires only two
values \cite{weymannPRB05}. As follows from Fig.~\ref{Fig:3}, TMR
in the linear response regime is independent of the gate voltage.
More precisely, it is given by $p^2/(1-p^2)$, which is exactly
equal to a half of the TMR calculated within the Julliere model
\cite{julliere75}. We recall that the Julliere value of TMR, ${\rm
TMR}_0=2p^2/(1-p^2)$, is characteristic of a single-barrier planar
tunnel junction. In the case of double-barrier planar junctions
with ferromagnetic leads, the TMR in the sequential transport
regime is equal to half of the Julliere value. This applies also
to sequential transport through quantum dots coupled to
ferromagnetic leads. However, one should bear in mind that
transport in the Coulomb blockade regime is dominated by
higher-order tunneling, which can considerably modify the
corresponding TMR. For instance, it was shown recently for
single-level quantum dots that TMR in the cotunneling regime
exhibits a strong dependence on the number of electrons on the
dot, and for empty or fully occupied dots reaches the Julliere's
value \cite{weymannPRB05}.

\begin{figure}[t]
\centering
  \includegraphics[width=0.45\columnwidth]{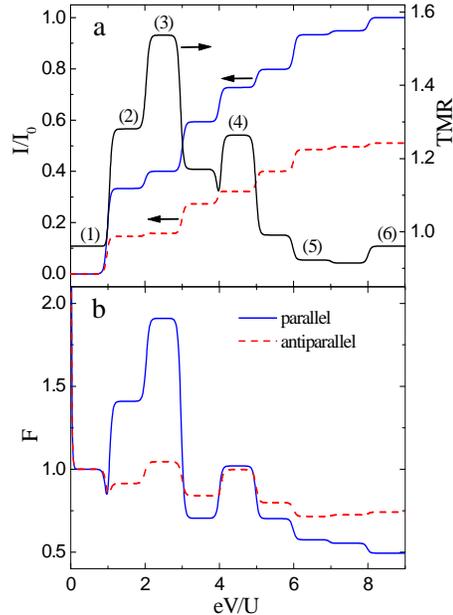}
  \caption{\label{Fig:4} (color online)
  The current (a) in units of $I_0=e\Gamma/\hbar$
  and Fano factor (b) in the
  parallel (solid line) and antiparallel
  (dashed line) configurations
  as well as tunnel magnetoresistance (a) as a function
  of the bias voltage for $\varepsilon=U/2$. The
  other parameters are the same as in Fig.~\ref{Fig:2}.}
\end{figure}

\begin{figure}[t]
\centering
  \includegraphics[width=0.45\columnwidth]{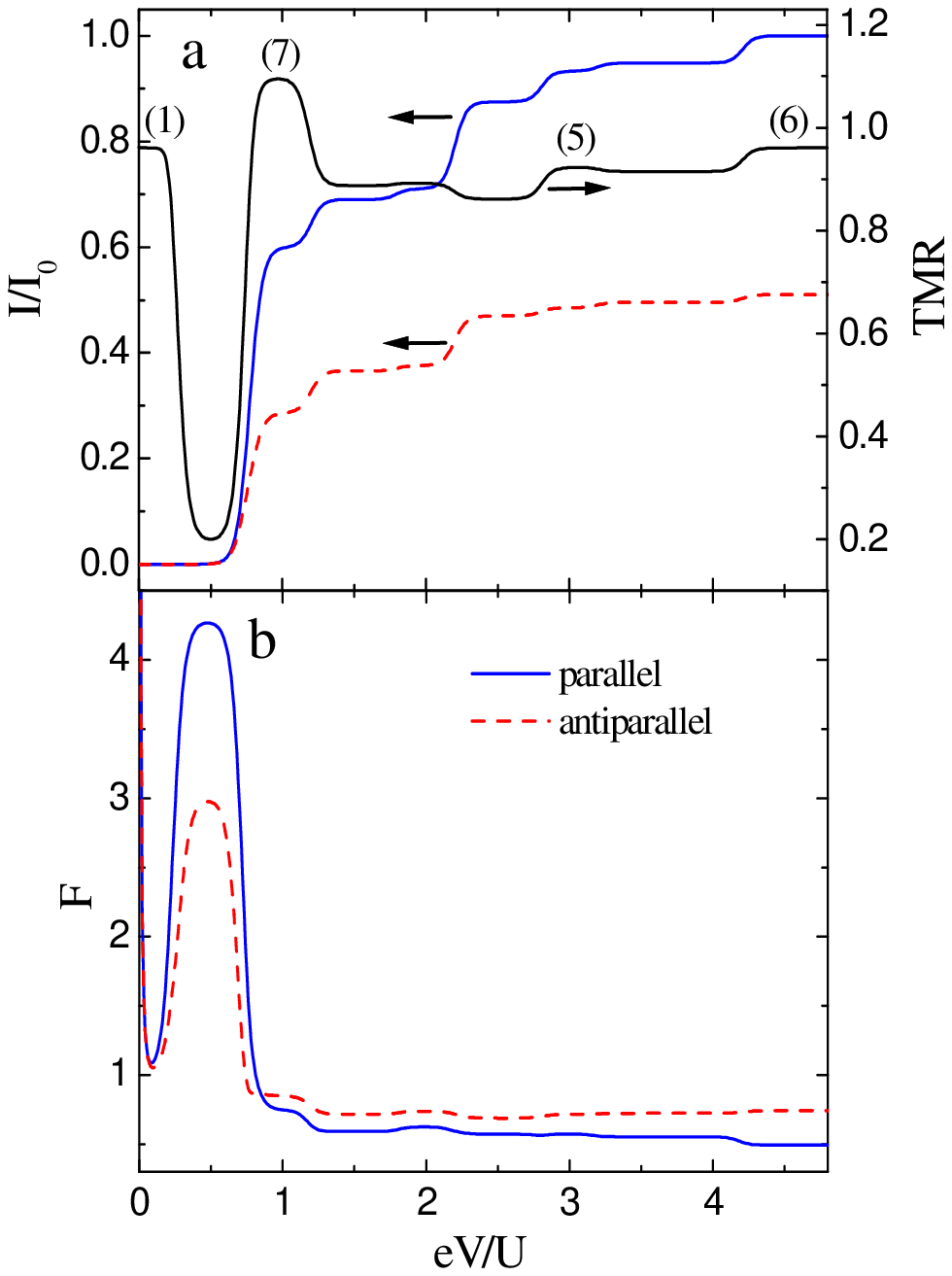}
  \caption{\label{Fig:5} (color online)
  The bias dependence of the current (a),
  the Fano factor (b) in both magnetic configurations
  and the TMR (a) calculated for $\varepsilon=-7U/5$.
  The other parameters are the same as in Fig.~\ref{Fig:4}.
  Current is plotted in units of $I_0=e\Gamma/\hbar$.}
\end{figure}

Figure~\ref{Fig:3} also shows that, when increasing the bias
voltage $V$ and keeping constant position of the dot levels, TMR
acquires some specific and well-defined values in different
transport regions. To analyze this behavior in more details, we
show in Figs.~\ref{Fig:4}(a) and \ref{Fig:5}(a) the bias voltage
dependence of the current and TMR for the case when the dot level
is above ($\varepsilon = U/2$) and below ($\varepsilon = -7U/5$)
the Fermi level of the leads at equilibrium. As one can see, the
current in both magnetic configurations and the associated TMR
exhibit characteristic plateaus which correspond to different
transport regions. When assuming the zero-temperature limit, one
can derive approximate formulas describing transport in each
region. However, the analytical solution in a general case is
rather complicated due to sixteen different eigenstates involved.
Therefore, in the following we only present the analytical results
for the transport regions, where the corresponding formulas are
relatively simple and transparent. These transport regions are
marked in Figs.~\ref{Fig:4}(a) and ~\ref{Fig:5}(a) by consecutive
numbers. An exemplary analytical calculation of the current in
parallel configuration for plateau (3) is described in the
appendix.

The analytical expressions for the current $I_{\rm P (AP)}$ and
TMR are listed in Table~\ref{tab:formulas}. First of all, the
current depends on the spin polarization of the leads effectively
only in the antiparallel configuration, while in the parallel
configuration it is independent of $p$. Such a behavior is a
consequence of the left-right symmetry for each spin channel in
the parallel configuration, and equal occupation of the charge
states active in transport. This leads to the independence of the
average charge and spin of the dot on the polarization factor $p$,
which in turn leads to the charge current independent of $p$. This
is not the case in the antiparallel configuration, where the
symmetry is absent and the current does depend on the spin
polarization of the leads. The associated TMR disappears for $p=0$
and diverges as $p\rightarrow 1$ (this is because $I_{\rm
AP}\rightarrow 0$ for $p\rightarrow 1$). For a finite spin
polarization $p$, one finds a considerable enhancement of TMR in
the regions (2), (3), and (4) as compared to the other transport
regions. The maximum TMR occurs in the region (3) and is given by
$4/5\times {\rm TMR}_0$. This enhancement can be accounted for by
realizing that in this transport region all the one-particle
states take part in transport, leading to an enhancement of the
spin accumulation in the antiparallel configuration. The doubly
occupied states do not take part in transport due to the Coulomb
energy. In the antiparallel configuration, spin-down electrons of
the source electrode can tunnel relatively easily to the dot,
where, however, they have to spend a longer time before tunneling
further to the drain electrode. Consequently, the current is
mainly determined by the tunneling probability through the barrier
between the dot and the drain electrode. This means that a new
channel for tunneling, which becomes active when going from the
region (2) to the region (3), has no significant influence on the
current, as it is clearly visible in Fig.~\ref{Fig:4}(a). In turn,
in the parallel configuration this new transport channel
contributes to the current, increasing this way the difference
between the parallel and antiparallel configurations (and
consequently the associated TMR).

\begin{table}
\caption{\label{tab:formulas}Analytical expressions approximating
the current $I_{\rm P}$ ($I_{\rm AP}$), the Fano factor $F_{\rm
P}$ ($F_{\rm AP}$) in the parallel (antiparallel) magnetic
configuration, and the TMR in different transport regimes
corresponding to plateaus shown in Figs.~\ref{Fig:4} and
~\ref{Fig:5} for $p_L=p_R\equiv p$. The current is expressed in
the units of $I_0=e\Gamma/\hbar$. In the Coulomb blockade regime,
where the current is suppressed, the TMR was determined from the
linear conductance.}
\begin{tabular}{cccccc}
Plateau & $I_{\rm P}$ & $I_{\rm AP}$ & TMR & $F_{\rm P}$ & $F_{\rm
AP}$ \\
\hline
\\
(1) & $0$ & $0$ & $\frac{p^2}{1-p^2}$ & $1$ & $1$ \\
\\
(2) & $\frac{1}{3}$ & $\frac{1-p^2}{3+p^2}$ &
$\frac{4p^2}{3-3p^2}$ &  $\frac{5+3p^2}{9(1-p^2)}$ &
$\frac{(5-p^2)(1+3p^2)}{(3+p^2)^2}$ \\
\\
(3) & $\frac{2}{5}$ & $\frac{2(1-p^2)}{5+3p^2}$ &
$\frac{8p^2}{5-5p^2}$ & $\frac{17+15p^2}{25(1-p^2)}$ &
$\frac{17+(62-15p^2)p^2}{(5+3p^2)^2}$ \\
\\
(4) & $\frac{8}{11}$ &  $\frac{8(1-p^2)}{11+(2+3p^2)p^2}$ &
$\frac{(13+3p^2)p^2}{11(1-p^2)}$ & $\approx\frac{0.55}{1-p^2}$ &
$\approx\frac{0.55+(2-p^2)(0.6+p^2)p^2}{(1+0.2p^2+0.3p^4)^2}$ \\
\\
(5) & $\frac{14}{15}$ & $\frac{2(1-p^2)(7+p^2)}{(5-p^2)(3+p^2)}$ &
$\frac{8(13+p^2)p^2}{15(1-p^2)(7+p^2)}$ &
$\approx\frac{0.5-0.45p^2}{1-p^2}$ &
$\approx\frac{5 + 7p^2}{10+4p^2}$ \\
\\
(6) & $1$ & $1-p^2$ & $\frac{p^2}{1-p^2}$ & $\frac{123+5p}{256}$ &
$\frac{123+(5+133p-5p^2)p}{256}$ \\
\\
(7) & $\frac{3}{5}$ & $\frac{(1-p^2)(3+p^2)}{5+(2+p^2)p^2}$ &
$\frac{8(2+p^2)p^2}{5(1-p^2)(3+p^2)}$ &
$\approx\frac{0.47-0.2p^2}{1-p^2}$ &
$\approx\frac{5(7+19p^2+26p^4)}{(3+p^2)(5+2p^2+p^4)^2}$ \\
\\
\end{tabular}
\end{table}

Although the current flowing through the system in the parallel
configuration is independent of the spin polarization in transport
regions marked by the numbers, we note that $I_{\rm P}$ slightly
depends on $p$ in the other regions. This is mainly due to the
fact that in these regions the occupations of the many-body dot
states being active in transport are not equal, leading to a small
spin accumulation even in the parallel configuration. For example,
in the region (2) there are 3 states taking part in transport,
with the occupation probabilities equal to 1/3; in the region (4)
there are 11 states (one empty, four single-particle and six
two-particle states),  all equally occupied with the probability
1/11, etc. In these regions, the charge current does not depend on
spin polarization. However, there are also regions where the
occupation probabilities depend on the spin polarization of the
leads, such as for example the region between (3) and (4). In this
region 10 dot states participate in transport [the same as in
region (4) except for the state with doubly occupied second
level]. We find that in this region the dot states do not take
part in transport on an equal footing, which leads to small spin
accumulation. This is a consequence of the interplay between the
energy structure of the dot and ferromagnetism of the electrodes.
The spin accumulation disappears when transport goes through
single-particle states. This simply follows from the same spin
asymmetry of tunneling through the left and right barriers. If we
have transport through doubly occupied states with Coulomb
interaction, the situation becomes more complex. The Coulomb
interaction in some situations may suppress this symmetry, leading
to spin accumulation. We note that similar spin accumulation for
parallel magnetic configuration in symmetric junctions was also
observed in different systems, see eg. \cite{barnasPRB00}.

A new and interesting behavior of TMR may be observed in the
blockade region, where the dot is doubly occupied, see
Figs.~\ref{Fig:3} and \ref{Fig:5}(a). Contrary to the case of a
single-level quantum dot, where sequential TMR is constant in the
whole blockade regime, the TMR displays now a strong dependence on
the bias voltage. Once the voltage is increased to around
$eV\approx U/2$, TMR drops rapidly from $1/2\times {\rm TMR}_0$ to
about $1/10 \times {\rm TMR}_0$ and then, when further increasing
the bias voltage, TMR increases to the value corresponding to the
plateau (7), see Tab.~\ref{tab:formulas}. This leads to a deep
minimum in TMR clearly visible in Fig.~\ref{Fig:5}(a), which can
be also seen in Fig.~\ref{Fig:3} as the two black areas in the
middle of the figure.

We have also calculated the Fano factor $F_{\rm P}$ and $F_{\rm
AP}$ in the parallel and antiparallel magnetic configurations,
respectively. The zero-frequency shot noise can be then found from
the knowledge of the current $I$ and the Fano factor $F$. The Fano
factor for the two situations discussed above, and corresponding
to Figs.~\ref{Fig:4}(a) and \ref{Fig:5}(a), is shown in
Figs.~\ref{Fig:4}(b) and \ref{Fig:5}(b) for both magnetic
configurations. Similarly to the case of current and TMR, the Fano
factor acquires roughly constant values, different in different
transport regions. The relevant analytical formulas, obtained with
the same assumptions as before, are given in
Table~\ref{tab:formulas}. Due to the spin asymmetry in the
coupling of the dot to external leads, the bias dependence of the
Fano factor is significantly different from that in the
corresponding nonmagnetic situations \cite{thielmann05}. For both
magnetic configurations of the system, the Fano factor depends on
the polarization factor $p$ (differently in the two
configurations, in general), except for the case of empty dot
where the exponentially suppressed transport gives rise to
Poissonian Fano factor $F_{\rm P}=F_{\rm AP}=1$. In addition, if
$|eV|\ll k_{\rm B}T$, the Fano factor becomes divergent due to the
thermal noise, which dominates in this transport regime; in the
case of $V=0$, the noise is given by $S = 4k_{\rm B}T G^{\rm
lin}$, with $G^{\rm lin}$ being the linear conductance, leading to
a divergency of the Fano factor (finite $S$ for $I = 0$)
\cite{blanterPR00,sukhorukovPRB01}. Furthermore, in some transport
regions we find $F_{\rm P}>F_{\rm AP}$, while in the other ones
$F_{\rm P} < F_{\rm AP}$. As one can see from the expressions
listed in Table~\ref{tab:formulas}, the ratio $F_{\rm P}/F_{\rm
AP}$ depends on the spin polarization of the leads $p$. For
example, in the region (2) $F_{\rm P}>F_{\rm AP}$ only if
$p\gtrsim 0.37$. Furthermore, if the leads are half-metallic, the
Fano factor in the parallel configuration diverges as
$p\rightarrow 1$, except for regions (1) and (6), where $F_{\rm
P}=1$ and $F_{\rm P}=1/2$, respectively. On the other hand, in the
antiparallel configuration the Fano factor tends to unity for
$p\rightarrow 1$, except for the Coulomb blockade regime with two
electrons trapped in the dot, where we find super-Poissonian shot
noise, $F_{\rm AP}\approx 3$, see Fig.~\ref{Fig:5}(b).

The super-Poissonian shot noise in the Coulomb blockade regime
with two electrons trapped in the dot, shown in
Fig.~\ref{Fig:5}(b), can be accounted for as follows. In the
ground state the dot is occupied by two electrons on the first
level (of lower energy). Assume the voltage corresponding to the
maximum of the peak in the Fano factor, $eV/U \approx 0.5$. The
system is still in the blockade regime [compare
Fig.~\ref{Fig:5}(a)], where the current is exponentially
suppressed. A nonzero small current can flow due to thermal
excitations. There is an exponentially small probability that one
electron leaves the dot and the other one jumps to the dot either
to the same (first) or to the second energy level (of higher
energy). If it tunnels to the second energy level, then transport
through this level is allowed and electron can easily leave the
dot, while another one can jump to the same level or to the level
of lower energy. If it tunnels to the same level, further
tunneling processes are allowed. If it tunnels to the low energy
(first) level, the system is blocked again. This leads to large
fluctuations in the current, and consequently to super-Poissonian
shot noise shown in Fig.~\ref{Fig:5}(b). This behavior does not
results from magnetism of the electrodes and persists even in the
case of nonmagnetic leads, where the Fano factor is approximately
$F\approx 3$. The effect is only quantitatively modified by
ferromagnetism of the electrodes -- the magnitude of the
super-Poissonian shot noise is different in parallel and
antiparallel configurations, as shown in Fig.~\ref{Fig:5}(b).

On the other hand, the increase of the Fano factor in the parallel
configuration, observed when $p\rightarrow 1$, is due to the
enhanced spin asymmetry in transport processes through the dot
\cite{bulka00,cottet04}. Consider for instance the region (2) in
Fig.~\ref{Fig:4}(b). Spin-up electrons tunnel then relatively
easily to and out of the dot. However, when a spin-down electron
appears on the dot, it blocks transport for relatively long time,
leading to large fluctuations in the current. These fluctuations
increase when $p\rightarrow 1$, giving rise to super-Poissonian
Fano factor, which is characterized by a divergent component
$F_{\rm P}\sim (1-p^2)^{-1}$. In turn, such fluctuations are
absent in the antiparallel configuration and the shot noise is
Poissonian, $F_{\rm AP}\sim 1$ for $p\rightarrow 1$.

\subsection{Quantum dots asymmetrically coupled to the leads}

In the previous subsections we have analyzed transport through
quantum dots coupled symmetrically to ferromagnetic leads (equal
spin polarizations of the leads and equal coupling parameters). In
that case transport characteristics were symmetric with respect to
the bias reversal. However, when the dot is coupled asymmetrically
to the leads, the $I-V$ curves are no longer symmetric with
respect to the bias reversal, leading to further interesting
effects. In the following, we discuss transport properties of
quantum dots asymmetrically coupled to the leads and consider two
special cases. The first case concerns the quantum dot coupled to
one nonmagnetic and one half-metallic lead ($p_{\rm L} \equiv p$,
$p_{\rm R} = 0$ and $\Gamma_{1r} = \Gamma_{2r^\prime}$), whereas
in the second case the dot is coupled to the leads with equal spin
polarizations but with unequal coupling strengths ($p_{\rm L} =
p_{\rm R} \equiv p$ and $\Gamma_{1r} \neq \Gamma_{2r^\prime}$). It
has been shown recently that transport characteristics of a
single-level quantum dots coupled to one half-metallic and one
nonmagnetic lead display a diode-like behavior
\cite{rudzinski01,bulka00,weymannPRB06}. On the other hand, in the
nonmagnetic limit of the second case a negative differential
conductance has been found \cite{thielmann05}.

\subsubsection{Unequal spin polarizations of the leads}

\begin{figure}[t]
\centering
  \includegraphics[width=0.45\columnwidth]{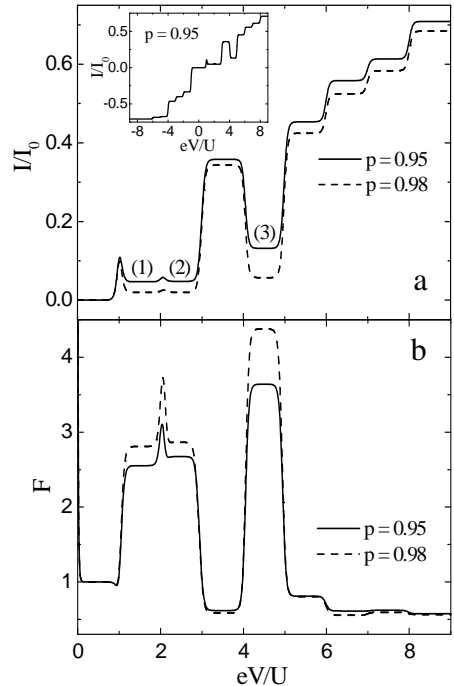}
  \caption{\label{Fig:6}
  The current (a) in units of $I_0=e\Gamma/\hbar$
  and Fano factor (b)
  as a function of the bias voltage for
  $p_{\rm L} \equiv p = 0.95, 0.98$, $p_{\rm R}=0$, while the
  other parameters are
  the same as in Fig.~\ref{Fig:2}. The inset
  in part (a) shows the current in the whole range of
  the bias voltage.}
\end{figure}

When one of the leads is half-metallic ($p=1$) and the other one
is nonmagnetic ($p=0$), transport characteristics become
asymmetric with respect to the bias reversal. Furthermore, the
current can be suppressed in certain  bias regions, and this
suppression is accompanied by the occurrence of NDC. This
basically happens when the electrons residing in the dot have spin
opposite to that of electrons in the half-metallic drain
electrode. In Fig.~\ref{Fig:6} we show the current and Fano factor
for a quantum dot coupled to half-metallic (left) and nonmagnetic
(right) lead as a function of the bias voltage. For the parameters
assumed, the current is suppressed in certain ranges of positive
bias voltage. As follows from Fig.~\ref{Fig:6}, there are three
blockade regions. To facilitate the further discussion, we label
these blockade regions with the numbers, as indicated in
Fig.~\ref{Fig:6}(a). On the other hand, for negative bias voltage,
the current changes monotonically with the transport voltage, as
one can see in the inset of Fig.~\ref{Fig:6}(a).

In the blockade region (1), $1\lesssim eV/U\lesssim 2$, the dot is
in the state $\ket{\downarrow}\ket{0}$ and the spin-down electron
residing in the dot has no possibility to tunnel further to the
left lead, which leads to suppression of the current. The blockade
region (2), $2\lesssim eV/U\lesssim 3$, is associated with the
occupation of the states $\ket{\downarrow}\ket{0}$ and
$\ket{0}\ket{\downarrow}$. The current is then prohibited due to
the full occupation of the single-particle spin-down states. When
increasing the bias voltage further, $3\lesssim eV/U\lesssim 4$,
the blockade of the current becomes suppressed [see the plateau
between regions (2) and (3) in Fig.~\ref{Fig:6}(a)], which is due
to a finite occupation of state $\ket{\uparrow\downarrow}\ket{0}$.
Although tunneling of spin-down electrons is then blocked, the
current is still carried by spin-up electrons. In turn, the
blockade region (3) occurs for $4\lesssim eV/U\lesssim 5$, where
the dot is in the triplet state $\ket{\downarrow}\ket{\downarrow}$
and tunneling is also suppressed. Thus, the current is blocked
when the total dot spin $S_z$ is either $S_z=-1/2$ or $S_z=-1$,
i.e., the spin of electrons on the dot is opposite to that of
electrons in the half-metallic lead. There is no suppression of
the current for negative voltage. This is because the electrons
residing in the dot can always tunnel to the nonmagnetic drain
electrode.

The current blockade in the regions (1) to (3) in
Fig.~\ref{Fig:6}(a) for positive bias is not due to the charging
effects as in the Coulomb blockade regime, but due to a particular
occupation of the dot spin state. Such blockade is frequently
referred to as the Pauli spin blockade, and has already been found
in single-dot and double-dot systems
\cite{weinmann95,fransson05,ono02}. Here, we show that the Pauli
spin blockade can occur in single two-level quantum dots, provided
they are coupled to half-metallic lead(s). It is further worth
noting that the blockade of the current in Fig.~\ref{Fig:6}(a) is
not complete -- there is a small leakage current inside each
blockade region, which results from the fact that the assumed spin
polarization of the half-metallic lead is not exactly equal to
unity. When spin polarization is increased, the current in the
blockade regions decreases (compare the curves for $p=0.95$ and
$p=0.98$).

The spin blockade of charge current leads to the super-Poissonian
shot noise, i.e., the corresponding Fano factor is larger than
unity. On the other hand, the Fano factor outside the spin
blockade regions is sub-Poissonian (smaller than unity), as shown
in Fig.~\ref{Fig:6}(b). The enhancement of the shot noise in the
Pauli blockade regions is a consequence of large spin asymmetry in
the tunneling processes. The occurrence of a spin-down electron on
the dot prevents further tunneling processes for a longer time,
while spin-up electrons on the dot can escape much faster,
allowing further tunneling processes. This gives rise to large
current fluctuations, and consequently also to Fano factors much
larger than unity.

Assuming the zero temperature limit, one can find approximate
expressions for the current (in the units of $I_0=e\Gamma/\hbar$)
in the spin blockade regions
\begin{equation}
  I^{(1)} = \frac{1-p^2}{3-p^2}  \,,
\end{equation}
for the first (1),
\begin{equation}
  I^{(2)} = \frac{2(1-p^2)}{5-p^2} \,,
\end{equation}
second (2), and
\begin{equation}
  I^{(3)} = \frac{ 2(1-p^2)(2-p^2) }{11 - 8p^2 + p^4} \,,
\end{equation}
for the third (3) blockade region, see Fig.~\ref{Fig:6}(a). The
above formulas show that the current vanishes when $p\rightarrow
1$. This is due to the total blockade of charge transport by
spin-down electrons in the dot. This can be also concluded from
the respective occupation probabilities, which  are given by
$P_{\ket{\downarrow}\ket{0}}^{(1)} = (1+p)/(3-p^2)$ for the first
blockade regime, $P_{\ket{\downarrow}\ket{0}}^{(2)} =
P_{\ket{0}\ket{\downarrow}}^{(2)} = (1+p)/(5-p^2)$ for the second
one, and $P_{\ket{\downarrow}\ket{\downarrow}}^{(3)} = (1+p)^2 /
(11-8p^2+p^4)$ for the third blockade region. It is clearly
evident from these formulas that the corresponding occupations
approach unity for $p\rightarrow 1$, leading to the full spin
blockade. It is also interesting to note that in region (3) one
observes formation of a pure triplet state.

With the same approximations as above, one can find the
corresponding formulas for the Fano factor in the spin blockade
regions
\begin{equation}
  F^{(1)} = \frac{(1+p^2)(5+p^2)}{(3-p^2)^2} \,,
\end{equation}
for the region (1), and
\begin{equation}
  F^{(2)} = \frac{17+30p^2+p^4}{(5-p^2)^2} \,,
\end{equation}
for the region (2). From the above expressions follows that
$F^{(1)}=5/9$ and $F^{(2)}=17/25$ for $p=0$, i.e., the shot noise
is sub-Poissonian, However, for $p\rightarrow 1$ the noise becomes
super-Poissonian, $F^{(1)}=3$ and $F^{(2)}=51/16$. In the third
spin blockade regime, the analytical formula for the Fano factor
$F^{(3)}$ is too cumbersome to be presented here. It is however
worth noting that $F^{(3)}\approx 0.55$ for $p=0$, while
$F^{(3)}=5$ for $p\rightarrow 1$. In addition, from the analytical
expressions for the Fano factor one can easily estimate the value
of spin polarization $p$ at which the shot noise becomes
super-Poissonian. This gives $F^{(1)}>1$ for $p>1/\sqrt{3}$, and
$F^{(2)}>1$ for $p>1/\sqrt{5}$.

\subsubsection{Unequal coupling of the dot levels to the leads }

\begin{figure}[t]
\centering
  \includegraphics[width=0.4\columnwidth]{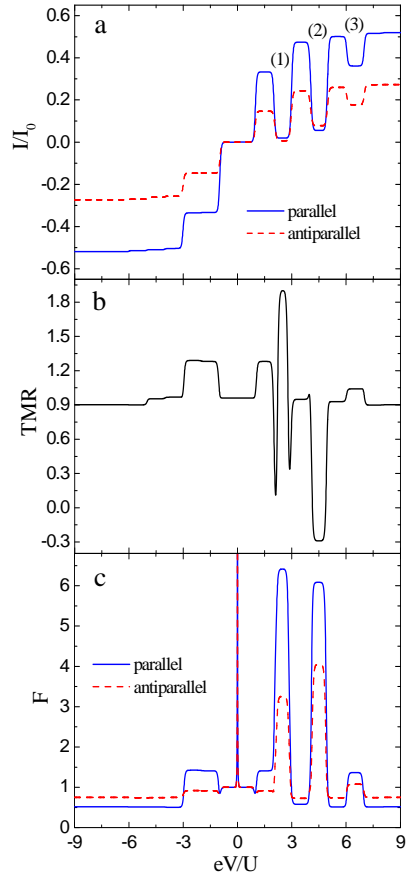}
  \caption{\label{Fig:7}
  (color online)
  The current (a) in units of $I_0=e\Gamma/\hbar$
  and the Fano factor (b) in the parallel
  (solid line) and antiparallel (dashed line) configurations, and
  the resulting TMR (c) as a function of the bias voltage for
  $\varepsilon = 25\Gamma$ and
  $\Gamma_{\rm L1} = \Gamma_{\rm R1} = 50\Gamma_{\rm L2} =
  \Gamma_{\rm R2} \equiv \Gamma/2$.
  The other parameters are
  the same as in Fig.~\ref{Fig:2}.}
\end{figure}

It has been shown recently that a negative differential
conductance (NDC) can occur in transport characteristics of
two-level quantum dots, when the two orbital levels are coupled to
nonmagnetic leads with different coupling strengths
\cite{thielmann05}. In the following, we consider the dot coupled
to ferromagnetic leads of equal spin polarizations, but with
unequal coupling strengths. For one bias polarization we find
narrow bias regions where NDC occurs, followed by the regions
where the current is partly suppressed. The suppression of current
is accompanied by a super-Poissonian shot noise and may lead to
negative TMR, as shown in Fig.~\ref{Fig:7}, where the current and
Fano factor in the parallel and antiparallel magnetic
configurations as well as the corresponding TMR are plotted as a
function of the bias voltage. As one can see, NDC and the
associated regions where charge current is partly suppressed occur
in both magnetic configurations of the system. The suppression
regions are marked with (1), (2) and (3), see Fig.~\ref{Fig:7}(a),
and occur for $2\lesssim eV/U \lesssim 3$, $4 \lesssim eV/U
\lesssim 5$, and $6 \lesssim eV/U \lesssim 7$, respectively.

In the blockade region (1), the dot can be occupied by at most one
electron. The blocking mechanism is associated with an increased
occupation of the second orbital level and decreased occupation of
the first level. Since the coupling of the second level to the
left lead (drain for electrons at positive bias) is the weakest
coupling in the system (the others are assumed to be equal), the
average occupation of this level is larger than the occupation of
the first level. Thus, an electron that has tunneled from the
right lead to the second level spends a long time in the dot
before it tunnels further to the right lead and, consequently,
blocks transport through the first level. In other words, the
onset of electron tunneling to the second level of the dot leads
to the suppression of current (NDC and the associated suppression
region). In turn, no NDC and current suppression takes place for
negative bias voltage, as now an electron that has tunneled to the
second level can easily tunnel further to the drain (right)
electrode. From the above discussion follows that the suppression
of current for positive bias voltage occurs in those transport
regions, where the occupation of the second level is larger than
that of the first dot level, and the rate for tunneling between
the second level and the drain electrode is smaller than that for
the source electrode. The same scenario also holds for the next
two blockade regions. In the region (2) it is the increased
occupation of the state $\ket{0}\ket{\uparrow \downarrow}$ which
is responsible for the blockade, whereas in the region (3)
transport is mainly mediated through the states
$\ket{0}\ket{\uparrow \downarrow}$, $\ket{\uparrow}\ket{\uparrow
\downarrow}$, and $\ket{\downarrow}\ket{\uparrow \downarrow}$. If
the second level is completely decoupled from the left lead,
transport in the region (3) is still mediated through the
single-particle states of the first orbital level. This is why
there is only a weak suppression of the current in this blockade
region, as compared to the first and second blockade regions,
where the current can be fully suppressed.

In the zero temperature limit one can derive some analytical
formulas for the current, TMR and Fano factor in all the three
blockade regions. First of all, it is interesting to note that NDC
in the regime (1) occurs when $x<1/3$ and $x<(1+p^2)/(3+p^2)$ for
the parallel and antiparallel configurations, respectively, with
$x=\Gamma_{\rm L2}/\Gamma$. The parameter $x$ describes the
asymmetry between the coupling of the second dot level to the left
lead and the other couplings, $\Gamma_{\rm L1}=\Gamma_{\rm
R1}=\Gamma_{\rm R2}=\Gamma/2$. Thus, for $x\rightarrow 0$ the
second dot level is completely decoupled from the left lead,
whereas for $x = 1/2$ one recovers the symmetric case. The current
(in the units of $I_0$) in both magnetic configurations is given
by $I_{\rm P}^{(1)} = 2x/(1+3x)$ and $I_{\rm AP}^{(1)} =
2(1-p^2)x/[1+3x+(1+x)p^2]$, whereas the TMR is
\begin{equation}
  {\rm TMR}^{(1)} = \frac{2p^2}{1-p^2}\frac{1+2x}{1+3x} \,.
\end{equation}
It is worth noting that the current tends to zero for
$x\rightarrow 0$, whereas TMR approaches ${\rm TMR}_0$. The
formulas for the Fano factor are too complex to be written here
explicitly. However, we find that for $x\rightarrow 0$, $F_{\rm
P}^{(1)} = (3+p^2)/(1-p^2)$ and $F_{\rm AP}^{(1)} =
(3+10p^2-p^4)/(1+p^2)^2$ for the parallel and antiparallel
configurations, respectively, which yields super-Poissonian shot
noise, irrespective of spin polarization of the leads.

It is also possible to derive analytical formula for the current,
TMR, and the conditions for NDC in the other blockade  regions.
However, they  are not simple and would make the discussion
somehow obscure. Therefore, we will only consider two limiting
situations, i.e., the symmetric case and the case of $x\rightarrow
0$, where the formulas become relatively simple. When the
couplings are symmetric, the blockade is lifted and the current,
TMR and Fano factor in the blockade regimes are given by the
expressions corresponding to the third, forth and fifth plateaus,
as listed in Table~\ref{tab:formulas}. When the second level is
decoupled from the left lead ($x\rightarrow 0$), current in the
second blockade regime tends to zero, while the TMR becomes
negative and is given by ${\rm TMR}^{(2)} = -4p^2/(3+p^2)$, see
Fig.~\ref{Fig:7}(b).

To understand the physical mechanism leading to the negative sign
of TMR let us assume a small value of $x$ and that the dot is in
the blocked state with two electrons on the second level. When a
spin-up electron tunnels from the dot to the left lead (this
tunneling probability is higher than for spin-down electrons), the
system becomes unblocked, and the next blockade takes place when
spin-up electron from the right lead tunnels to the dot. This
tunneling probability is larger in the parallel configuration than
in the antiparallel (in the former case it involves spin majority
electrons while in the latter case spin minority ones). Thus, the
blockade is more effective in the parallel configuration than in
the antiparallel one, leading to negative TMR for small values of
$x$, as shown in Fig.~\ref{Fig:7}(b). For the parameters assumed
to calculate Fig.~\ref{Fig:7}, the TMR changes sign when
$x\lesssim 0.023$. For the Fano factor in the limit of
$x\rightarrow 0$ we find $F_{\rm AP}^{(2)} > F_{\rm P}^{(2)}$ for
$p<1$. However, if $p\rightarrow 1$ the Fano factor is divergent
in both magnetic configurations of the system. Furthermore, in the
limit of $x\rightarrow 0$, the shot noise becomes
super-Poissonian, irrespective of magnetic configuration of the
system and spin polarization $p$. For $p=0$, the Fano factor is
minimum and given by $F_{\rm P}^{(2)} = F_{\rm AP}^{(2)} \approx
4.9$.

Contrary to the first two blockade regions, current in the third
blockade region is finite for $x\rightarrow 0$. This is due to the
fact that although tunneling through the second level is
suppressed, current can be still mediated by single-particle
states of the first level. In the zero temperature limit the
current in the parallel and antiparallel configurations is given
by (in the units of $I_0$) $I^{(3)}_{\rm P}=1/3$ and $I_{\rm
AP}^{(3)} = (1-p^2)/(3+p^2)$, respectively. This yields the tunnel
magnetoresistance ${\rm TMR}^{(3)} = 4p^2 / (3-3p^2)= 2/3\times
{\rm TMR}_0$, which is characteristic for the transport regime
where single-particle states take part in transport, see the
formulas for plateau (2) in Table~\ref{tab:formulas}. The Fano
factor in the case of negligible coupling $\Gamma_{\rm L2}$ in
both magnetic configurations is $F_{\rm P}^{(3)} = (5+3p^2) /
(9-9p^2)$ and $F_{\rm AP}^{(3)} = (5-p^2)(1+3p^2) / (3+p^2)^2$,
respectively. This implies that $F_{\rm AP}^{(3)}\leq 1$ for all
$p$, while $F_{\rm P}^{(3)}$ becomes larger than unity for
$p>1/\sqrt{3}$.

As follows from the above discussion, a distinctively different
transport behavior of the system can be found in each blockade
region, when the coupling of the second level to the left lead is
negligible. This is especially visible in TMR which in the first
blockade region approaches the Julliere's value, in the second
blockade changes sign and becomes negative, while in the third
blockade region is given by $2/3$ of the Julliere value.
Furthermore, in the two first blockade regions we find a full
suppression of the current for $x \rightarrow 0$, accompanied by a
super-Poissonian shot noise. It is also worth noting, that the
origin of the blockade is now different from that discussed
previously (Pauli spin blockade), although some qualitative
features of the blockade are very similar.

\subsection{Effect of external magnetic field}

So far we have assumed that the dot levels are spin degenerate.
However, in a strong magnetic field the Zeeman energy can be large
enough (of the order of $\Gamma$ or larger) to influence transport
characteristics significantly \cite{kogan04}. The impact of an
external magnetic field on transport properties is the most
visible in the case of symmetric couplings, therefore only this
situation is discussed in the following.

\begin{figure}[t]
\centering
  \includegraphics[width=0.4\columnwidth]{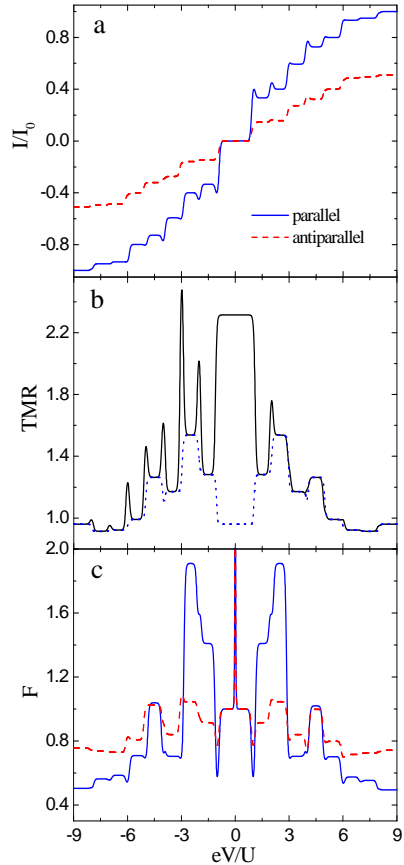}
  \caption{\label{Fig:8}
  (color online)
  The current (a) in units of $I_0=e\Gamma/\hbar$
  and Fano factor (c) in the parallel
  (solid line) and antiparallel (dashed line) configurations
  as well as the TMR (b) as a function of the bias voltage
  in the presence of external magnetic field, $\Delta=5\Gamma$.
  The other parameters are
  the same as in Fig.~\ref{Fig:2}. For comparison
  we also show the TMR in the case of $\Delta=0$,
  see the dotted blue line in part (b).}
\end{figure}

Bias dependence of the current and Fano factor in both magnetic
configurations and the corresponding TMR are shown in
Fig.~\ref{Fig:8} for the symmetric system and in the presence of
external magnetic field. Because of a finite Zeeman splitting of
the dot levels, there is now an asymmetry of transport
characteristics with respect to the bias reversal. Moreover, NDC
occurs now for both positive and negative bias voltages,
irrespective of magnetic configuration of the system, as clearly
visible in Fig.~\ref{Fig:8}(a). The NDC is however more pronounced
in the parallel configuration than in the antiparallel one.

The TMR is displayed in Fig.~\ref{Fig:8}(b), where, for
comparison, the TMR in the absence of magnetic field is also
shown. If the energy levels are split due to the Zeeman energy,
TMR exhibits characteristic peaks which occur at voltages
corresponding to the Coulomb steps in the I-V curves. These peaks
are more visible for negative bias voltage. In addition, a high
plateau evolves in the low bias range due to the splitting of the
dot levels. To determine the analytic formula for the TMR in the
linear response, we consider the linear conductance, which in the
case of $\Delta \gg k_{\rm B}T$ and low temperature can be
approximated by $G^{\rm lin}_{\rm P}\sim (1+p)/2$ and $G^{\rm
lin}_{\rm AP}\sim (1-p^2)/2$, for the parallel and antiparallel
configurations, respectively. This yields for the TMR
\begin{equation}
  {\rm TMR}^{\rm lin}=p/(1-p) \,.
\end{equation}
It is worth noting that TMR is now much enhanced as compared to
the case of no external magnetic field, where the TMR is given by
$p^2/(1-p^2)$. Moreover, the linear TMR in the presence of
magnetic field is even larger than the Julliere value of TMR. The
enhancement of TMR is due to the fact that when $\Delta \gg k_{\rm
B}T$, the thermally-activated sequential transport takes place
only through the state $\ket{\uparrow}\ket{0}$ and not through
states $\ket{\uparrow}\ket{0}$ and $\ket{\downarrow}\ket{0}$ as in
the case of $\Delta=0$. This considerably increases the spin
asymmetry of the total current in both magnetic configurations,
and gives rise to enhanced TMR in the linear response regime. The
peak structure of TMR results from the corresponding peaks in
current, shown in Fig.~\ref{Fig:8}(a). More specifically, when the
dot levels are spin-split, the two spin contributions to current
differ not only in magnitude, but also in positions of the steps
(peaks in the corresponding differential conductance). Adding the
two contributions leads to characteristics of the current with a
pronounced peak structure. This structure is much more visible in
the parallel configuration than in the antiparallel one. This
simply follows from spin asymmetry -- note that for vanishing
Zeeman splitting the two contributions to current become
equivalent in the antiparallel configuration, but are still
significantly different in the parallel one.

The Fano factor in both magnetic configurations is shown in
Fig.~\ref{Fig:8}(c). Contrary to the current and TMR, behavior of
the Fano factor is only slightly changed as compared to the case
of quantum dot in the absence of magnetic field, see
Figs.~\ref{Fig:4} and \ref{Fig:8}(c). This is because Zeeman
splitting does not provide a mechanism leading to a qualitative
change of the noise - the noise and current are modified in a
similar way, so the corresponding  Fano factor is only weakly
affected by magnetic field.

\section{Summary}

We have considered spin-polarized transport through a two-level
quantum dot coupled to ferromagnetic leads in the sequential
tunneling regime. The cases of symmetric and asymmetric coupling
of the dot to external leads have been analyzed. For the most
characteristic transport regions we have derived approximate
analytical formulas for the current and Fano factor in the
parallel and antiparallel configurations, as well as for the TMR
effect. These formulas may be useful in the interpretation of
future experimental data.

Transport properties of quantum dots symmetrically coupled to the
leads are shown to be strongly dependent on the bias and gate
voltages. First of all, the bias and gate voltage dependence of
TMR is then significantly different from that in the case of
single-level quantum dots. The TMR can take now several
well-defined values, which depend on the parameters of the system.
Furthermore, we have shown that the shot noise in the parallel
configuration may become super-Poissonian in some transport
regions. In the antiparallel configuration the noise is rather
sub-Poissonian and the Fano factor becomes equal to unity when the
leads become half-metallic.

We have also shown that for a quantum dot coupled to one
half-metallic and one nonmagnetic lead, the current is suppressed
in certain bias regions. The suppression of the current results
from the full occupation of the relevant dot spin state. A
super-Poissonian shot noise has been found in these blockade
regions. On the other hand, in the case when there is an asymmetry
between the couplings of the dot levels to external leads, some
blockade regions accompanied by NDC have been observed in both
magnetic configurations of the system. We found three blockade
regions with significantly different behavior of the TMR effect.
In the regions where NDC occurs, the Fano factor is larger than
unity and shot noise is super-Poissonian.

Additionally, the effect of a finite Zeeman splitting of the dot
levels was discussed in the case of quantum dot coupled
symmetrically to the leads. We have shown that NDC occurs in both
magnetic configurations, and for both bias polarizations.
Furthermore, TMR exhibits peaks at voltages corresponding to
consecutive Coulomb steps, and a high plateau in TMR appears in
the linear response regime. Unlike the current and TMR, the Fano
factor was found to be only slightly affected by an external
magnetic field.

The numerical results presented in this paper have been calculated
for $U_1=U_2=U^\prime$. However, the results for a general case,
where the Coulomb integral between two electrons localized on
different orbitals is different from the Coulomb integral for two
electrons occupying the same orbital are qualitatively similar and
the difference is rather of quantitative nature. More
specifically, the lengths of different plateaus on the voltage
scale may now be different, which is a consequence of the fact
that the charging energy depends now on the distribution of
electrons between the two orbitals of the quantum dot. Apart from
this, the heights of the plateaus may also be different as
compared to the case when all Coulomb integrals are equal.


\ack

We acknowledge discussions with J. Fransson, J. K\"onig, J.
Martinek, R. \'Swirkowicz, and M. Wilczy\'nski. This project was
supported by funds of the Polish Ministry of Science and Higher
Education as a research project in years 2006-2009. I. W.
acknowledges support from the Foundation for Polish Science.

\appendix

\section*{Appendix}

In the following we describe the procedure to derive the
analytical formulas presented in table \ref{tab:formulas}. As an
example, we calculate the current in the parallel configuration
for the transport region (3) shown in fig. \ref{Fig:4}. In this
region there are five dot states taking part in transport, these
are the empty state and all the four single-particle states, see
table \ref{tab:states}. As the occupations and contribution to the
current coming from the other states are suppressed, it is
sufficient to consider only the transitions (self-energies)
between the above-mentioned five states. In the low temperature
limit we approximate the Fermi-Dirac functions by step functions
and get for the full self-energy matrix in the basis defined by
these five dot states
\begin{equation*}
\fl  \mathbf{W} = \left(\begin{array}{ccccc}
    -2\Gamma & (1+p)\Gamma/2 & (1-p)\Gamma/2 & (1+p)\Gamma/2 & (1-p)\Gamma/2\\
    (1+p)\Gamma/2 & -(1+p)\Gamma/2 & 0 & 0 & 0\\
    (1-p)\Gamma/2 & 0 & (p-1)\Gamma/2 & 0 & 0\\
    (1+p)\Gamma/2  & 0 & 0 & -(1+p)\Gamma/2 & 0\\
    (1-p)\Gamma/2 & 0 & 0 & 0 & (1-p)\Gamma/2\\
  \end{array} \right)
\end{equation*}
By replacing one row of $\mathbf{W}$ with $(\Gamma,\dots,\Gamma)$,
which is due to normalization, one can calculate the occupation
probabilities from Eq. (\ref{Eq:master}). For plateau (3) and
parallel configuration one finds that the probabilities are equal
and given by $1/5$, i.e. each of the five states is equally
occupied. In a similar way one can determine the elements of
self-energy matrix $\mathbf{W^I}$
\begin{equation*}
\fl  \mathbf{W^I} = \left(\begin{array}{ccccc}
    0 & -(1+p)\Gamma/2 & (p-1)\Gamma/2 & -(1+p)\Gamma/2 & (p-1)\Gamma/2\\
    -(1+p)\Gamma/2 & 0 & 0 & 0 & 0\\
    (p-1)\Gamma/2 & 0 & 0 & 0 & 0\\
    -(1+p)\Gamma/2  & 0 & 0 & 0 & 0\\
    (p-1)\Gamma/2 & 0 & 0 & 0 & 0\\
  \end{array} \right)
\end{equation*}
It is now possible to calculate the current from Eq.
(\ref{Eq:current}), which gives $I_{\rm P} = 2e\Gamma/5\hbar$. The
other analytical expressions can be determined in the same manner.


\end{document}